\documentclass[12pt]{article}

\usepackage{amsmath}
\usepackage{amssymb}
\usepackage{amsfonts}
\usepackage{latexsym}
\usepackage{color}

\catcode `\@=11 \@addtoreset{equation}{section}

\catcode `\@=12

\newcommand{\be}{\begin{equation}}
\newcommand{\en}{\end{equation}}
\newcommand{\bea}{\begin{eqnarray}}
\newcommand{\ena}{\end{eqnarray}}
\newcommand{\beano}{\begin{eqnarray*}}
\newcommand{\enano}{\end{eqnarray*}}
\newcommand{\bee}{\begin{enumerate}}
\newcommand{\ene}{\end{enumerate}}

\newcommand{\N}{\mathfrak N}

\newcommand{\mc}{\mathcal}

\newcommand{\F}{{\cal F}}

\newcommand{\Lc}{{\cal L}}

\newcommand{\1}{1 \!\! 1}

\newcommand{\Hil}{\mc H}

\catcode `\@=11 \@addtoreset{equation}{section}
\catcode `\@=12

\begin{document}

\thispagestyle{empty}

\vspace*{2cm}

\begin{center}
{\Large \bf  Dissipation evidence for the quantum damped harmonic oscillator via pseudo-bosons}\\[10mm]

{\large F. Bagarello} \footnote{ Dipartimento di Metodi e Modelli Matematici, Facolt\`a di Ingegneria, Universit\`a di Palermo, I-90128
Palermo, ITALY. \, e-mail: bagarell@unipa.it\,\,\,\, Home page: www.unipa.it/$^\sim$bagarell}
\vspace{3mm}\\

\end{center}

\vspace*{2cm}

\begin{abstract}
\noindent It is known that a self-adjoint, time-independent hamiltonian can be defined for the quantum damped harmonic oscillator. We show here that the two vacua naturally
associated to this operator, when expressed in terms of pseudo-bosonic lowering and raising operators, appear to be non square-integrable. This
fact is interpreted as the evidence of the dissipation effect of the classical oscillator at a purely quantum level.

\end{abstract}

\vspace{2cm}

%{\bf PACS Numbers}:  .......

\vfill

%\pagenumbering{roman}

\newpage

\section{Introduction}

In a recent paper, \cite{abg}, we have analyzed the quantum damped harmonic oscillator within the context of pseudo-bosonic operators, which
were introduced recently by Trifonov in \cite{tri} and further analyzed in a series of papers by us, \cite{baglast}. The strategy adopted in \cite{abg} to quantize the system was that proposed in \cite{ban} following an idea given in \cite{fesh}, idea which goes back to Bateman, \cite{bat}, and which proved to be the most natural  for our purposes. However, it is well known that Bateman's idea is not the only possibility of quantizing a generic dissipative system, and the damped quantum harmonic oscillator in particular: many different approaches exist and we refer to \cite{chru} for a detailed list of references. We just recall here that in quantum mechanics a dissipative system is quite often seen as an open system interacting with some reservoir, so that the energy of the system needs not be preserved, in general, and the unitary group of automorphisms which represents the dynamics for closed systems is replaced by a certain dynamical semigroup.

Going back to our results in \cite{abg}, we have shown that, even
if pseudo-bosonic operators naturally appear in the analysis of the model, the vacua of the two annihilation operators of the model cannot
belong to the Hilbert space $\Hil=\Lc^2({\Bbb R}^2)$ in which the system is assumed to live. Hence the general structure proposed in
\cite{baglast} apparently does not work. However, this is not the end of the story, since the details of our strategy are not at all uniquely determined. Thus it is
natural to wonder whether different choices exist which allow us to construct the same functional settings as in \cite{baglast}. In this paper
we will continue this analysis, answering to this question.

The paper is organized as follows:  in the next section we introduce and discuss two-dimensional pseudo-bosons analyzing some of their
mathematical properties.  In Section III we introduce and analyze the quantum damped harmonic oscillator, while Section IV contains our
conclusions.

\section{The pseudo-bosonic settings}

Let $\Hil$ be a given Hilbert space with scalar product $\left<.,.\right>$ and related norm $\|.\|$. We introduce two pairs of operators, $a_j$
and $b_j$, $j=1,2$, acting on $\Hil$ and satisfying the following commutation rules \be [a_j,b_j]=\1, \quad \mbox{ and }\quad
[a_1,a_2]=[a_1,b_2]=[a_2,b_1]=0. \label{21} \en  Of course, they collapse to the CCR's for the independent modes if $b_j=a^\dagger_j$, $j=1,2$.
It is well known that $a_j$ and $b_j$ are unbounded operators, so they cannot be defined on all of $\Hil$. Following \cite{baglast}, and
writing $D^\infty(X):=\cap_{p\geq0}D(X^p)$ (the common  domain of all the powers of the operator $X$), we consider the following:

\vspace{2mm}

{\bf Assumption 1.--} there exists a non-zero $\varphi_{0,0}\in\Hil$ such that $a_j\varphi_{0,0}=0$, $j=1,2$, and $\varphi_{0,0}\in
D^\infty(b_1)\cap D^\infty(b_2)$.

{\bf Assumption 2.--} there exists a non-zero $\Psi_{0,0}\in\Hil$ such that $b_j^\dagger\Psi_{0,0}=0$, $j=1,2$, and $\Psi_{0,0}\in
D^\infty(a_1^\dagger)\cap D^\infty(a_2^\dagger)$.

\vspace{2mm}

Under these assumptions we can introduce the following vectors in $\Hil$: \be
\varphi_{n,l}=\frac{1}{\sqrt{n!\,l!}}\,b_1^n\,b_2^l\,\varphi_{0,0} \quad \mbox{ and }\quad
\Psi_{n,l}=\frac{1}{\sqrt{n!\,l!}}(a_1^\dagger)^n(a_2^\dagger)^l\Psi_{0,0}, \quad n,l\geq 0. \label{22}\en Let us now define the unbounded
operators $N_j:=b_ja_j$ and $\N_j:=N_j^\dagger=a_j^\dagger b_j^\dagger$, $j=1,2$.  It is possible to check that $\varphi_{n,l}$ belongs to the
domain of $N_j$, $D(N_j)$, and $\Psi_{n,l}\in D(\N_j)$, for all $n,l\geq0$ and for $j=1,2$. Moreover, \be N_1\varphi_{n,l}=n\varphi_{n,l},
\quad N_2\varphi_{n,l}=l\varphi_{n,l}, \quad \N_1\Psi_{n,l}=n\Psi_{n,l}, \quad \N_2\Psi_{n,l}=l\Psi_{n,l}. \label{23}\en

Under the above assumptions, if we chose the normalization of $\Psi_{0,0}$ and $\varphi_{0,0}$ in such a way that
$\left<\Psi_{0,0},\varphi_{0,0}\right>=1$, we get \be \left<\Psi_{n,l},\varphi_{m,k}\right>=\delta_{n,m}\delta_{l,k}, \quad \forall
n,m,l,k\geq0. \label{27}\en This means that the sets $\F_\Psi=\{\Psi_{n,l},\,n,l\geq0\}$ and $\F_\varphi=\{\varphi_{n,l},\,n,l\geq0\}$ are {\em
biorthogonal} and, because of this, the vectors of each set are linearly independent. We further assume that

\vspace{2mm}

{\bf Assumption 3.--} $\F_\Psi$ and $\F_\varphi$ are complete in $\Hil$.

\vspace{2mm}

Let us now introduce the operators $S_\varphi$ and $S_\Psi$ via their action respectively on the bases $\F_\Psi$ and $\F_\varphi$: \be
S_\varphi\Psi_{n,k}=\varphi_{n,k},\qquad S_\Psi\varphi_{n,k}=\Psi_{n,k}, \label{213}\en for all $n, k\geq0$. These imply that
$\Psi_{n,k}=(S_\Psi\,S_\varphi)\Psi_{n,k}$ and $\varphi_{n,k}=(S_\varphi \,S_\Psi)\varphi_{n,k}$, for all $n,k\geq0$. Hence \be
S_\Psi\,S_\varphi=S_\varphi\,S_\Psi=\1 \quad \Rightarrow \quad S_\Psi=S_\varphi^{-1}. \label{214}\en In other words, both $S_\Psi$ and
$S_\varphi$ are invertible and one is the inverse of the other. Furthermore, we can also check that they are both positive, well defined and
symmetric, \cite{baglast}. Moreover, at least formally, it is possible to write these operators in the bra-ket notation as \be
S_\varphi=\sum_{n,k=0}^\infty |\varphi_{n,k}><\varphi_{n,k}|,\qquad S_\Psi=\sum_{n,k=0}^\infty |\Psi_{n,k}><\Psi_{n,k}|. \label{212}\en
 These expressions are
only formal, at this stage, since the series may not converge in the uniform topology and the operators $S_\varphi$ and $S_\Psi$ could be
unbounded. This aspect was exhaustively discussed in \cite{baglast}, together with many other features of pseudo-bosons which are not relevant for us here.

It is interesting to observe that these two-dimensional pseudo-bosons give rise to interesting intertwining relations among non self-adjoint
operators, see \cite{bagpb3} and references therein. In particular it is easy to check that \be S_\Psi\,N_j=\N_jS_\Psi \quad \mbox{ and }\quad
N_j\,S_\varphi=S_\varphi\,\N_j, \label{219}\en $j=1,2$. This is related to the fact that the eigenvalues of, say, $N_1$ and $\N_1$ coincide and
that their eigenvectors are related by the operators $S_\varphi$ and $S_\Psi$.

\section{Quantum damped harmonic oscillator}

In \cite{abg} we have considered the quantum damped harmonic oscillator (QDHO) in connection with pseudo-bosons. Since the procedure is highly
non unique, the negative results we have obtained in \cite{abg}  only suggest the fact that {\em something peculiar} may happen. More
explicitly, we have shown that the vacua of the pseudo-bosonic annihilation operators do not belong to $\Lc^2({\Bbb R}^2)$. This fact,
however, leaves open the possibility that other choices are more appropriate than the ones considered in \cite{abg}, and in particular that the
choice of the Hilbert space where the model should be considered was not the most appropriate.

In this section we briefly review what we have done in \cite{abg} and then we propose different alternative approaches. All these proposals,
however, give rise to the same conclusion: it seems to be impossible to have an Hilbert space in which the two vacua of the pseudo-bosonic
annihilation operators live! We will comment on this result at the end of this section.

\subsection{Working in $\Lc^2({\Bbb R}^2)$}

The original equation of motion of a classical damped oscillator, \cite{fesh}, $m\ddot x+\gamma \dot x+kx=0$, is complemented by a second {\em
virtual} equation,  $m\ddot y-\gamma \dot y+ky=0$, and the classical lagrangian for the system looks like $L=m\dot x\dot
y+\frac{\gamma}{2}(x\dot y-\dot xy)-kxy$. This corresponds to a classical Hamiltonian $H=p_x\,\dot x+p_y\,\dot
y-L=\frac{1}{m}\left(p_x+\gamma\frac{y}{2}\right)\left(p_y-\gamma\frac{x}{2}\right)+kxy$, where $p_x=\frac{\partial L}{\partial\dot x}$ and
$p_y=\frac{\partial L}{\partial\dot y}$ are the conjugate momenta. The introduction of pseudo-bosons is based on two successive changes of
variables and on a canonical quantization. First of all we introduce the new variables $x_1$ and $x_2$ via $x=\frac{1}{\sqrt{2}}(x_1+x_2)$,
$y=\frac{1}{\sqrt{2}}(x_1-x_2)$. Then $L=\frac{1}{2}m\left(\dot x_1^2-\dot x_2^2\right)+\frac{\gamma}{2}\left(x_2\dot x_1-x_1\dot
x_2\right)-\frac{k}{2}(x_1^2-x_2^2)$ and
$H=\frac{1}{2m}\left(p_1-\gamma\frac{x_2}{2}\right)^2+\frac{1}{2m}\left(p_2+\gamma\frac{x_1}{2}\right)^2+ \frac{k}{2}(x_1^2-x_2^2)$. The second
change of variable is the following:

\bea
 \left\{
    \begin{array}{ll}
p_+=\sqrt{\frac{\omega_+}{2m\Omega}}p_1+i\,\sqrt{\frac{m\Omega\omega_+}{2}}\,x_2,\\
p_-=\sqrt{\frac{\omega_-}{2m\Omega}}p_1-i\,\sqrt{\frac{m\Omega\omega_-}{2}}\,x_2,\\
x_+=\sqrt{\frac{m\Omega}{2\omega_+}}x_1+i\,\sqrt{\frac{1}{2m\Omega\omega_+}}\,p_2,\\
x_-=\sqrt{\frac{m\Omega}{2\omega_-}}x_1-i\,\sqrt{\frac{1}{2m\Omega\omega_-}}\,p_2,\\
      \end{array}
        \right.\label{51} \ena
where we have introduced $\Omega=\sqrt{\frac{1}{m}\left(k-\frac{\gamma^2}{4m}\right)}$ and the  two  complex quantities $\omega_\pm=\Omega\pm
i\frac{\gamma}{2m}$. In the rest of the paper we will assume that $k\geq \frac{\gamma^2}{4m}$, so that $\Omega$ is real. Up to now, we are
still at a classical level, so that $\overline\omega_+=\omega_-$, $\overline p_+=p_-$, $\overline x_+=x_-$, and consequently, see below,
$\overline H_+=H_-$ and $\overline H=H$. Hence $H$ is a real Hamiltonian. Indeed, with these definitions, the Hamiltonian looks like the
hamiltonian of a two-dimensional harmonic oscillator
$$
H=\frac{1}{2}\left(p_+^2+\omega_+^2x_+^2\right)+\frac{1}{2}\left(p_-^2+\omega_-^2x_-^2\right)=:H_++H_-
$$
at least formally.

At this stage we quantize canonically  the system. Following \cite{ban}, we require that the following commutators are satisfied: \be
[x_+,p_+]=[x_-,p_-]=i\1, \label{52}\en  the other commutators being trivial. We also have to require that $ p_+^\dagger=p_-$ and that
$x_+^\dagger=x_-$, which are the quantum version of the {\em compatibility} conditions above.  The pseudo-bosons now appear:
 \bea
 \left\{
    \begin{array}{ll}
a_+=\sqrt{\frac{\omega_+}{2}}\left(x_++i\,\frac{p_+}{\omega_+}\right),\\
a_-=\sqrt{\frac{\omega_-}{2}}\left(x_-+i\,\frac{p_-}{\omega_-}\right),\\
b_+=\sqrt{\frac{\omega_+}{2}}\left(x_+-i\,\frac{p_+}{\omega_+}\right),\\
b_-=\sqrt{\frac{\omega_-}{2}}\left(x_--i\,\frac{p_-}{\omega_-}\right),\\
      \end{array}
        \right.\label{53} \ena
and indeed we have $[a_+,b_+]=[a_-,b_-]=\1$. Notice also that $b_+=a_-^\dagger$ and $b_-=a_+^\dagger$.
Moreover $H$ can be written in term of the operators $N_\pm=b_\pm a_\pm$ as $H=\omega_+N_++\omega_-N_-+\frac{\omega_++\omega_-}{2}\,\1$. So the
hamiltonian of the QDHO is quite simply written in terms of pseudo-bosonic operators.

In \cite{abg} we have used the following representation of the operators in (\ref{52}): \bea
 \left\{
    \begin{array}{ll}
x_+=\frac{1}{\Gamma\,\overline{\delta}-\delta\,\overline{\Gamma}}\left(\overline{\Gamma}\,p_y+\overline{\delta}\,x\right),\\
x_-=\frac{-1}{\Gamma\,\overline{\delta}-\delta\,\overline{\Gamma}}\left({\Gamma}\,p_y+{\delta}\,x\right),\\
p_+=\Gamma \,p_x+\delta \,y,\\
p_-=\overline{\Gamma} \,p_x+\overline{\delta} \,y,\\
      \end{array}
        \right.\label{54} \ena
for all fixed choices of $\Gamma$ and $\delta$ such that $\Gamma\,\overline{\delta}\neq \delta\,\overline{\Gamma}$. Here $x$, $y$, $p_x$ and
$p_y$ are pairwise conjugate self-adjoint operators: $[x,p_x]=[y,p_y]=i\1$. Notice that these operators also satisfy the compatibility
conditions $ p_+^\dagger=p_-$ and $x_+^\dagger=x_-$. Hence it is natural to represent $x$ and $y$ as the standard multiplication operators and
$p_x$ and $p_y$ as $-i\,\frac{\partial}{\partial\,x}=-i\,\partial_x$ and $-i\,\frac{\partial}{\partial\,y}=-i\,\partial_y$. Then we get

\bea
 \left\{
    \begin{array}{ll}
a_+=\sqrt{\frac{\omega_+}{2}}\,\left\{\left(\beta\,x+i\,\frac{\delta}{\omega_+}\,y\right)+\left(\frac{\Gamma}{\omega_+}\,
\partial_x-i\,\alpha\,\partial_y\right)\right\},\\
a_-=\sqrt{\frac{\omega_-}{2}}\,\left\{\left(\overline{\beta}\,x+i\,\frac{\overline{\delta}}{\omega_-}\,y\right)+
\left(\frac{\overline{\Gamma}}{\omega_-}\,\partial_x-i\,\overline{\alpha}\,\partial_y\right)\right\},\\
b_+=\sqrt{\frac{\omega_+}{2}}\,\left\{\left(\beta\,x-i\,\frac{\delta}{\omega_+}\,y\right)-\left(\frac{\Gamma}{\omega_+}\,
\partial_x+i\,\alpha\,\partial_y\right)\right\},\\
b_-=\sqrt{\frac{\omega_-}{2}}\,\left\{\left(\overline{\beta}\,x-i\,\frac{\overline{\delta}}{\omega_-}\,y\right)-
\left(\frac{\overline{\Gamma}}{\omega_-}\,\partial_x+i\,\overline{\alpha}\,\partial_y\right)\right\},\\
      \end{array}
        \right.\label{55} \ena
where, to simplify the notation, we have introduced $\alpha=\frac{\overline{\Gamma}}{\Gamma\,\overline{\delta}-\delta\,\overline{\Gamma}}$ and
$\beta=\frac{\overline{\delta}}{\Gamma\,\overline{\delta}-\delta\,\overline{\Gamma}}$. Notice that, since these operators satisfy the pseudo-bosonic commutation rules, the coefficients in (\ref{55}) satisfy the equalities $\alpha\,\overline{\delta}=\beta\,\overline{\Gamma}$ and $\beta\,{\Gamma}-\alpha\,{\delta}=1$.

Due to the fact that $b_+=a_-^\dagger$ and $b_-=a_+^\dagger$, Assumptions 1 and 2 of Section II collapse and we just have to look for a single
square-integrable function $\varphi_{0,0}(x,y)$ such that, first of all, $a_+\varphi_{0,0}(x,y)=a_-\varphi_{0,0}(x,y)=0$.
  It is possible to check that a solution of
$a_+\varphi_{0,0}(x,y)=a_-\varphi_{0,0}(x,y)=0$ is the following: \be
\varphi_{0,0}(x,y)=N_\varphi\,\exp\left\{-\,\frac{\beta\,\omega_+}{2\,\Gamma}\,x^2+\frac{\delta}{2\,\alpha\,\omega_+}\, y^2\right\},
\label{56}\en where it is required that the equality
$\frac{\omega_+}{\omega_-}=-\,\frac{\delta}{\overline{\delta}}\,\frac{\Gamma}{\overline{\Gamma}}$ holds, since in this way $\varphi_{0,0}(x,y)$
satisfies both $a_+\varphi_{0,0}(x,y)=0$ and $a_-\varphi_{0,0}(x,y)=0$. Of course, we would like $\varphi_{0,0}(x,y)$ to be square integrable. It is
easy to check that $\frac{\beta\,\omega_+}{2\Gamma}$ and $\frac{\delta}{\alpha\,\omega_+}$ are real. This follows from the above relation
between $\omega_+$ and $\omega_-$. Hence $\varphi_{0,0}(x,y)\in\Lc^2({\Bbb R}^2)$  only if $\frac{\beta\,\omega_+}{\Gamma}>0$ and if, at the
same time, $\frac{\delta}{\alpha\,\omega_+}<0$. However, using the previous relations between the coefficients, it follows that
$$
\frac{\beta\,\omega_+}{\Gamma}\frac{\delta}{\alpha\,\omega_+}=\frac{\delta\,\beta}{\alpha\,\gamma}=
\left|\frac{\delta}{\gamma}\right|^2>0.
$$
Hence $\varphi_{0,0}(x,y)$ cannot belong to $\Lc^2({\Bbb R}^2)$. We conclude that
Assumptions 1 and 2 of Section II are not satisfied, so that apparently there is no possibility of constructing two biorthogonal bases of
$\Lc^2({\Bbb R}^2)$ out of the QDHO.

\subsection{Changing Hilbert space}

This result does not exclude, however, that a solution of Assumptions 1 and 2 could be found in a different $\Lc^2$ space, for instance in a space with
a suitable weight. A simple-minded idea would be to replace $\Lc^2({\Bbb R}^2)$ with, for instance, $\Hil_1:=\Lc^2({\Bbb
R}^2,e^{-c_1x^2-c_2y^2}\,dx\,dy)$, where $c_1$ and $c_2$ should be two positive constants chosen in such a way that the wave-function
$\varphi_{0,0}(x,y)$ in (\ref{56}) does belong to $\Hil_1$. This approach, however, has an immediate drawback: in $\Hil_1$ the adjoint of the
operator is different from the one in $\Lc^2({\Bbb R}^2)$, and this should be taken into account to produce a consistent model. This is what we
will do in this section. In particular we will show that what we expected is reasonable but false!

In $\Hil_1$ the scalar product is clearly defined as follows: $\left<f,g\right>_1=\int_{\Bbb R}dx\int_{\Bbb
R}dy\,\overline{f(x)}\,g(x)\,e^{-c_1x^2-c_2y^2}$. The adjoint $X^*$ of the operator $X$ in $\Hil_1$,  is defined by the equation
$\left<Xf,g\right>_1=\left<f,X^*g\right>_1$, for all $f, g\in\Hil_1$ belonging respectively to the domain of $X$ and $X^*$. We refer to
\cite{rs} for further and rigorous reading on the definition of the adjoint for unbounded operators. It is easy to check that
$\partial_x^*=-\partial_x+2c_1x$ and $\partial_y^*=-\partial_y+2c_2y$, which return the adjoint in $\Lc^2({\Bbb R}^2)$ if $c_1=c_2=0$. Taking
as our starting point the operators $a_\pm$ and $b_\pm$ defined in (\ref{55}), we can compute their {\em new} adjoints (i.e. their adjoints in
$\Hil_1$), which can be written as \bea
 \left\{
    \begin{array}{ll}
a_+^*=b_-+\sqrt{2\omega_-}\left(c_1x\,\frac{\overline\Gamma}{\omega_-}+ic_2y\overline{\alpha}\right),\\
a_-^*=b_++\sqrt{2\omega_+}\left(c_1x\,\frac{\Gamma}{\omega_+}+ic_2y{\alpha}\right),\\
b_+^*=a_-+\sqrt{2\omega_-}\left(-c_1x\,\frac{\overline\Gamma}{\omega_-}+ic_2y\overline{\alpha}\right),\\
b_-^*=a_++\sqrt{2\omega_+}\left(-c_1x\,\frac{\overline\Gamma}{\omega_-}+ic_2y{\alpha}\right).\\
      \end{array}
        \right.\label{71} \ena
It is clear that these again reduce to the adjoints in $\Lc^2({\Bbb R}^2)$ when $c_1=c_2=0$. It is also clear that, but for this case, the
compatibility conditions required above are not satisfied: $a_\pm^*\neq b_\mp$. Nevertheless, if we carry on our analysis, we can still
look for the solutions of the differential equations $a_+\varphi_{0,0}(x,y)=a_-\varphi_{0,0}(x,y)=0$ and
$b_+^*\Psi_{0,0}(x,y)=b_-^*\Psi_{0,0}(x,y)=0$. The solution $\varphi_{0,0}(x,y)$ is, clearly, exactly the one in (\ref{56}), with the same
condition on the ratio $\frac{\omega_+}{\omega_-}$ as before. The wave-function $\varphi_{0,0}(x,y)$ belongs to $\Hil_1$ if the following
inequalities are satisfied: \be c_1+\frac{\beta\omega_+}{\Gamma}>0,\qquad c_2-\frac{\delta}{\alpha\omega_+}>0. \label{72}\en Notice also that $\varphi_{0,0}(x,y)$ is eigenvector of $H$ with eigenvalue $\frac{1}{2}\left(\omega_++\omega_-\right)$. This may appear as
an improvement with respect to the result of the previous section, since non trivial choices of $c_1$ and $c_2$ for which (\ref{72}) are
satisfied do exist. For any such choice $\varphi_{0,0}(x,y)$ belongs to $\Hil_1$ and it also belongs to the domain of all the powers of $b_-$
and $b_+$, so that the wave-functions $\varphi_{n_+,n_-}(x,y)$ can be defined as in (\ref{22}). Let us now look for the function
$\Psi_{0,0}(x,y)$. Due to (\ref{71}) we get \be
\Psi_{0,0}(x,y)=N_\Psi\,\exp\left\{-\,\frac{\beta\,\omega_+}{2\,\Gamma}\,x^2+\frac{\delta}{2\,\alpha\,\omega_+}\,
y^2\right\}\exp\left\{c_1x^2+c_2y^2\right\}, \label{73}\en which belongs to $\Hil_1$ if \be \frac{\beta\omega_+}{\Gamma}-c_1>0,\qquad
c_2+\frac{\delta}{\alpha\omega_+}<0. \label{74}\en  $\Psi_{0,0}(x,y)$ is eigenvector of $H^*$, which is different from $H^\dagger=H$, with eigenvalue $\frac{1}{2}\left(\omega_++\omega_-\right)$ It is not hard to check, now, that these conditions are not compatible with those in
(\ref{72}): in other words it is not possible to fix $c_1$ and $c_2$ in such a way both (\ref{72}) and (\ref{74}) are satisfied. This means
that our original simple-minded idea that adding a weight in the scalar product of the Hilbert space should regularize the situation does not
work as expected. More explicitly, if $c_1$ and $c_2$ satisfy (\ref{72}), then $\Psi_{0,0}(x,y)$ does not satisfy Assumption 2. Viceversa, if they satisfy
(\ref{74}), then $\varphi_{0,0}(x,y)$ does not satisfy Assumption 1. In both cases, therefore, only a single set of functions in $\Hil_1$ can
be constructed, which is (most likely) a basis of $\Hil_1$ itself.

\subsection{And now, what?}

Changing Hilbert space has proven an interesting exercise but it doesn't give positive results. For this reason, in this section, we stay in
the original Hilbert space and we consider four operators $a_\pm$ and $b_\pm$ satisfying $b_\pm=a_\mp^\dagger$ and the following commutation
rules, for convenience given only in terms of $a_\pm$ and of their adjoints: \be [a_+,a_-^\dagger]=\1,\quad [a_+,a_-]=[a_+,a_+^\dagger]=[a_-,a_-^\dagger]=0.
\label{75} \en A rather general expression for $a_\pm$, which extends the one in (\ref{55}), is the following \bea
 \left\{
    \begin{array}{ll}
a_+=\alpha_xx+\alpha_yy+\beta_x\partial_x+\beta_y\partial_y,\\
a_-=\gamma_xx+\gamma_yy+\eta_x\partial_x+\eta_y\partial_y,\\
      \end{array}
        \right.\label{76} \ena
where $\alpha_x$, $\alpha_y$, $\ldots$, $\eta_y$ are complex constants to be fixed. Due to the form of the operators, it is natural to look for solutions of
equations $a_\pm f_{0,0}(x,y)=0$ in the form $f_{0,0}(x,y)=N_0e^{-k_1x^2-k_2y^2}$, for some positive $k_1$ and $k_2$. This choice would
guarantee that $f_{0,0}(x,y)$ belongs not only to $\Lc^2({\Bbb R}^2)$, but also to $D^\infty(b_+)\cap D^\infty(b_-)$. However, it is easy to
check that, with these choices, no solution exist: simple algebraic manipulations show that $\beta_x=\beta_y=\eta_x=\eta_y=0$ and,
consequently, that also $\alpha_x=\alpha_y=\gamma_x=\gamma_y=0$, which is not possible. Hence, if we want to keep the form of the operators as
in (\ref{76}), we have to change the expression of $f_{0,0}(x,y)$. The simplest extension is $f_{0,0}(x,y)=N_0e^{-k_1x^2-k_2y^2-k_3xy}$, where
$k_3$ is a third real or complex constant to be fixed and which does not have to destroy the square-integrability of $f_{0,0}$. It is possible
to check that, if a solution does exist, it must correspond to non real values of the $\beta$'s and of the $\eta$'s. Indeed it is again a matter of simple algebraic
computation to check that, if $\beta_x$, $\beta_y$, $\eta_x$ and $\eta_y$ are assumed to be real, then no solution of the system does exist. It is also possible to
check that, if $k_3=2\sqrt{k_1\,k_2}$, no solution exist both for real and for complex values of the coefficients in (\ref{76}). In other words:
in all the situations which we have completely under control the solution simply does not exist, in the sense that we may have square
integrability in, say, $x$ but not in $y$ or vice-versa. Or, yet, we can have square integrability of $\varphi_{0,0}(x,y)$ in both $x$ and $y$
but not of the second wave-function $\Psi_{0,0}(x,y)$. It is also clear that changing the definitions in (\ref{76}) is dangerous since it can
destroy the compatibility conditions or the commutation rules, or both. Analogously, changing the expression for $f_{0,0}(x,y)$ could easily destroy the
validity of Assumption 1: our choices seems to be the only possible ones!

At a first sight these results may appear unpleasant. On the other hand, in our opinion they are very much physically motivated. Indeed, let us
recall our starting point: our {\em real} system, a damped harmonic oscillator, was coupled to a {\em virtual}, somehow forced, harmonic
oscillator. Hence the energy lost, at a classical level, from the first oscillator was transferred to the second, in such a way that the full
system stays conservative. This classical flux of energy produces, at a quantum level, the impossibility of having both $\varphi_{0,0}$ and
$\Psi_{0,0}$ in a single Hilbert space: for instance, if $\varphi_{0,0}$ can be normalized, $\Psi_{0,0}$ can not: its norm increases too much!
And this result (or the related ones) seems to be independent of several possible way out that one may imagine: if we change Hilbert space to
allow more functions in, since we also have to change  the adjoint,  we go out of the space anyway. Or we can look for other, and
possibly more convenient, representations of the pseudo-bosonic operators. But, once again, requiring that their vacua are square-integrable
produces a set of algebraic equations with no solution at all. Needless to say, our choices do not cover all the possible situations, so that,
in principle, other rather special solutions reflecting the construction in Section II could be found. However, we claim that these solution cannot exist, exactly because of the non conservative nature of our oscillator.

\section{Conclusions}

In this paper we have considered in more details the QDHO already discussed in \cite{abg}. We have shown that working in a different Hilbert
spaces or changing the representation of the pseudo-bosonic ladder operators do not change the main results of our previous analysis, i.e. that
the formal eigenstates of the number operators are not all square-integrable.  This impossibility has been interpreted here as intrinsically
related to the non conservative nature of the quantum system under analysis.

\section*{Acknowledgements}
   The authors would like to acknowledge financial support from the MIUR.

\end{document}